# Topological creation and destruction of edge states in photonic graphene

Mikael C. Rechtsman[1], Yonatan Plotnik[1], Julia M. Zeuner[2], Alexander Szameit[2], and Mordechai Segev[1]

[1]*Physics Department and Solid State Institute, Technion, Haifa 32000, Israel*
[2]*Institute of Applied Physics, Abbe Center of Photonics, Friedrich-Schiller-Universität Jena, Max-Wien-Platz 1, 07743 Jena, Germany*

[mcr@technion.ac.il](mcr@technion.ac.il)

*Abstract*

We demonstrate theoretically and experimentally a topological transition of classical light in 'photonic graphene': an array of waveguides arranged in the honeycomb geometry. As the system is uniaxially strained (compressed), the two unique Dirac points (present in the spectrum of conventional graphene) merge and annihilate each other, and a band gap forms. As a result, edge states are created on the zig-zag edge and destroyed on the bearded edge. These results are applicable for any 2D honeycomb-type structure, from carbon-based graphene to photonic lattices and crystals.

The rise of the first truly two-dimensional material, graphene, started with its discovery in 2004[1] and has continued as a result of its tremendous potential in electronic and optoelectronic devices, including applications in flexible electronics[2], optical modulation[3], and metamaterials[4], among others. Graphene has also been shown to be extremely important and unique in the fundamental understanding of condensed matter physics[5]. The key reason is the presence of two unique Dirac cones in graphene's band structure, leading to quasiparticle dynamics governed by the (dispersionless) Dirac equation, instead of the more typical Schrödinger equation; this causes electrons to propagate similarly to massless relativistic fermions. Indeed, a model of a graphene-like structure was used in proposing the first topological insulator[6-8], an insulating material with conducting edges whose propagating edge states are immune to scattering disorder. This immunity has led to active research into whether such edge states can remain entangled for long times, and act as robust qubits in a quantum computer[8]. The striking feature of topological physical properties is that they are not affected by small perturbations (at any order in perturbation theory) of the system. For example, edge states of two-dimensional topological insulators cannot backscatter unless the scattering potential breaks time-reversal symmetry or is large enough to close the bulk gap. Graphene, while not a topological insulator, possesses edge states that are 'topologically protected' in the sense that their presence is derived from a topological property (the Berry phase) and cannot be destroyed by weak perturbations[9-11]. Only for very strong perturbations can a 'topological transition' be observed: the Dirac points merge and a bulk band gap is opened[12].

The unique dispersion properties of graphene have been exploited in the field of photonics as well. The presence of photonic Dirac cones has been demonstrated experimentally[13] and studied theoretically in numerous papers[13–20]. More specifically, experiments with photonic graphene have thus far addressed conical diffraction[13], topological protection of edge states in magnetic photonic crystals[21], enabling single photonic mode behavior over large areas[22], and pseudomagnetic behavior at optical frequencies[23]. Systems exhibiting photonic Dirac cones include photonic lattices (waveguide arrays)[13,16,17,19], and two- and three-dimensional photonic crystals[22,24]. Edge states that can be derived from topological arguments have also been demonstrated optically in systems without Dirac cones, namely one-dimensional quasicrystalline waveguide arrays[25]. Light propagating in photonic lattices (arrays of optical waveguides) with honeycomb geometries obeys exactly the same dynamical equation as electrons in graphene, namely the Schrödinger equation with a honeycomb potential. Therefore, such "photonic graphene" lattice emulates carbon-based graphene in much the same way that Bose-Einstein condensates in optical lattices emulate condensed matter phenomena. Indeed, all of the topological properties of graphene carry over to honeycomb photonic lattices.

In this Letter, we experimentally and theoretically demonstrate a topological transition in photonic graphene. A photonic graphene lattice is compressed (or uniaxially strained), and undergoes a transition - as a function of the degree of compression - from an ungapped phase with two unique Dirac points, to a phase where the Dirac points have merged and a band gap opens. This transition is universal to any honeycomb-shaped potential in any wave system, yet the compression required in observing this transition is beyond the physically applicable strain levels in carbon-based graphene. Hence, photonic graphene provides an ideal setting for experimenting with this fundamental phenomenon. Since each of the Dirac points in the uncompressed graphene has a Berry phase of $\pi$ and -$\pi$, respectively, they are topologically protected against small strains or compressions that preserve inversion symmetry (which the compression indeed preserves). We show and experimentally demonstrate that when the compression reaches a certain critical threshold, the states localized on the edges of the photonic graphene undergo a major transition. Namely, the edge states associated with the so-called "bearded" (or "Klein") edge disappears when the Dirac points merge, while conversely, the edge states associated with the "zig-zag" edge then occupy the entire edge Brillouin zone. This behavior is associated with the merging of the Dirac points and the subsequent formation of a bulk band gap. The edge terminations are depicted graphically in Fig. 1(a). Thus, probing the confinement on the edges of the photonic graphene sample provides an ideal means for identifying the transition and studying the features associated with it.

The paraxial discrete Schrödinger equation describes the diffraction of light through a photonic lattice[26,27]:

$$i\partial_z \psi_n(z) = \sum_{\langle m \rangle} c(|\boldsymbol{r}_{n,m}|)\psi_m(z) \equiv H_{m,n}\psi_m. \tag{1}$$

Here, $z$ is the distance of propagation; $\psi_n$ is the amplitude of the guided mode in the $n^{th}$ waveguide; $c(|r_{m,n}|)$ is the coupling constant between waveguides $m$ and $n$ when they are placed a distance $|r_{m,n}|$ apart from one another; $H$ is the Hamiltonian matrix; and the summation is taken over only the nearest neighbor waveguides. Note that this equation is equivalent to a tight-binding model of electrons in a lattice, where the $z$-coordinate takes the place of the time coordinate of electron evolution. Thus as light diffracts through the lattice, it behaves analogously to electrons evolving in time in the solid state.

The waveguide array is arranged in a honeycomb lattice (i.e., similarly to graphene), as shown in Fig. 1(a) in a microscope image of the input facet. We fabricate a total of six samples, with an increasing degree of affine compression in the vertical direction (Fig. 1(b) shows the sample with the greatest compression). The vertical coupling constant, $c_1$, increases more rapidly with compression than the diagonal coupling constant, $c_2$, (see Fig. 1(c)). Note that while the waveguides are not isotropic, we show below that their anisotropy has little effect on the results. In the honeycomb lattice geometry, the Hamiltonian may be represented in momentum space as:

$H(k) = h(k) \cdot \sigma$, where

$$h(k) = \left(c_1 + 2c_2 \cos\left(\frac{k_x}{2}\right) \cos\left(\frac{\sqrt{3}k_y}{2}\right), 2c_2 \cos\left(\frac{k_x}{2}\right) \sin\left(\frac{\sqrt{3}k_y}{2}\right)\right) \qquad (2)$$

where $H(k)$ is the Hamiltonian represented in lattice momentum ($k = (k_x, k_y)$) space and $\sigma = (\sigma_x, \sigma_y)$ is a two-dimensional vector of the $x$ and $y$ Pauli matrices. Diagonalization of the Hamiltonian results in the band structure diagrams shown in Fig. 2(a) and Fig. 2(b), for the cases of the uncompressed honeycomb lattice ($c_1=c_2$) and the highly compressed lattice ($c_1=2.5c_2$). Note that the physical meaning of the eigenvalue of the Hamiltonian is the wavenumber in the z-direction (also known as the propagation constant), or $\beta(\mathbf{k})$. In the uncompressed case, the band structure exhibits Dirac cones (conical intersections of the two bands), two of which are inequivalent (all others are separated by a reciprocal lattice vector from the principal two). When the lattice is compressed, the Dirac cones move together (as indicated by the black arrows in Fig. 2(a))[12,17]. When the compression is such that $c_1=2c_2$, the cones merge, and a band gap is opened in the spectrum[17].

The presence of edge states can be analytically derived from the "bulk-edge correspondence[10,11,28–30]," via a calculation involving the Berry phase (or "Zak phase[31]") - a topological property of the band structure of the system – as explained here. The Berry phase can be written as:

$$\gamma = i\oint d\mathbf{k} \cdot \langle \phi_k | \nabla_k | \phi_k \rangle, \tag{3}$$

where the integral is over a closed loop in $\mathbf{k}$-space, and $|\phi_k\rangle$ is an eigenstate of the Hamiltonian (in either the first or second band) at Bloch wavevector $\mathbf{k}$. The eigenstate can be expressed as $|\phi_\mathbf{k}\rangle=(1,e^{i\theta_k})/2^{1/2}$, where $\theta_k$ is a real phase that is a function of the Bloch wavevector $\mathbf{k}$. It can be shown[31] that $\gamma$ is a *topological invariant* of a lattice system – i.e, that small changes in the nature of the shape of the loop cannot change the value of $\gamma$. Moreover, specifically in the honeycomb lattice $\gamma = w\pi$, where $w$ is the *winding number* of the vector $\mathbf{h}(\mathbf{k})$. Note that the angle that $\mathbf{h}(\mathbf{k})$ makes with the positive $k_x$-axis is exactly $\theta_k$, and Eq. (3) amounts to integrating $\theta_k/2$ along the given integration path. Thus, if along the path through $\mathbf{k}$-space the vector

$h(k)$ makes a complete loop, then $\gamma = \pi$, otherwise, $\gamma = 0$, but no other values. Furthermore, any compression of the honeycomb lattice does not change these values. In Fig. 3(a), we represent this calculation pictorially in $k$-space, with the Dirac points indicated by black dots. The path taken by the integral follows the green and red vertical lines, and the arrows represent the direction of the vector $h(k)$. Note that even though the integration paths are vertical lines, they are in fact closed loops because the Brillouin zone is periodic, and thus the path effectively connects with its beginning. The diagram is shown for a unit cell that results in the bearded edge termination, when it is repeated infinitely in the $x$-direction but is terminated in the $y$-direction. The bulk-edge correspondence theorem states that if $\gamma = \pi$, there will be an edge state at the value of $k_x$ corresponding to the vertical line, whereas if $\gamma = 0$, there will not be an edge state there. It is clear from the bold arrows shown in Fig. 3(a) that, in the shaded region, the vector $h(k)$ makes a complete loop and thus there are bearded edge states at these values of $k_x$, whereas in the unshaded regions there are not. The zig-zag edge contains edge states for values of $k_x$ for which the bearded edge contains none (unshaded regions), and does not contain edge states for which the bearded edge does (shaded regions).

As the lattice is compressed, the Dirac points get closer to one another in pairs, and then merge and open a band gap when $c_1=2c_2$[17]. Until they merge, they are topologically protected against opening a band gap due to the Berry phase associated with circular loops around them (see section 1 of the Appendix for more details on the nature of the topological protection of the Dirac points). Figure 3(b) shows an equivalent plot to that of Fig. 3(a), but for supercritical compression ($c_1=2.5c_2$). Here, the Dirac points have come together, entirely eliminating the bearded edge state (i.e., the shaded region of Fig. 3(a)). Conversely, the zig-zag edge then occupies the entire "edge Brillouin zone" (i.e., all values of $k_x$). The diagram equivalent to Fig. 3, but for the zig-zag edge, is shown in section 2 of the Appendix. To confirm the presence (and absence) of the edge states on the bearded and zig-zag edges, we explicitly compute the edge band structures for a number of different values of the compression, *s*. This is done by choosing a unit cell that is periodic in the x-direction, but which is terminated with alternately the bearded and zig-zag edges in the y-direction. The result of this band structure calculation (which uses the full Schrödinger equation with laboratory parameters) is shown and discussed in section 3 of the Appendix.

In our experiments, the honeycomb waveguide array is written using the femtosecond-direct-laser-writing technique[32], in fused silica. The waveguides are elliptical in shape, with horizontal diameters of 11μm and 3μm, respectively. The index of refraction of the ambient silica is 1.45, and the change of index associated with the waveguides is *6x 10^-4*. The nearest-neighbor spacing between waveguides is 22μm, making the lattice constant *a=22×√3μm*. To probe exactly where the edge state resides as a function of $k_x$, we perform a type of "spatial spectroscopy" on the edge, as follows. An elliptical beam of light (at wavelength *633nm*, from a Helium-Neon laser) is launched at the input facet of the array, such that it is localized at the edge, but is broad in the direction parallel to the edge. The broadness of the beam in the *x*-direction implies that is narrow in $k_x$ space and thus excites states in a narrow line width around $k_x$. The elliptical beam is then tilted in order to tune the value of $k_x$, thus applying a linear phase gradient in the *x*-direction. The precise details of the experimental setup are described in detail in Ref. [33].

Next, we present the numerical and experimental results associated edge confinement, starting first with the bearded edge (the top of the waveguide array depicted in Fig. 1(a)). The numerical results are obtained using the beam propagation method[34], in which the continuous Schrödinger equation is evolved in "time" (in propagation distance, *z*) in a finite honeycomb waveguide array. Figures 4(a) and 4(b) show (simulation and experiment, respectively) the ratio of optical power remaining on the edge (within the first two rows of waveguides) relative to the power diffracted into the bulk, plotted as a function of Bloch wavenumber, $k_x$, within the edge Brillouin zone. In each plot, results are presented for arrays compressed in the vertical direction by a factor of *s = 0.5, 0.6, 0.7, 0.8, 0.9,* and *1.0*, yielding coupling constant ratios of $c_1/c_2$ = 3.6, 2.6, 1.9, 1.5, 1.1, and 0.9, respectively. Note that the ratio of the vertical coupling, $c_1$, to the diagonal coupling, $c_2$, is at its greatest when the lattice is at its most compressed state (*s = 0.5*). For *s = 1.0*, the uncompressed array, there is strong edge confinement surrounding $k_x$ = *0*, indicating the presence of the edge state. This is precisely the edge state derived in the discussion on Fig. 2. The strong edge confinement near $k_x = \pi/a$ is due to the presence of another, non-topological edge state that has been previously demonstrated elsewhere[33], and is not examined in the present work. With increasing compression, the results in Fig. 4 show progressively lower power remaining on the edge, as well as a narrower region in $k_x$ -space for which power is confined there. This corresponds directly to the destruction of the bearded edge state demonstrated in Fig. 2. The fraction of power remaining on the edge also decreases as a result of the fact that the edge states penetrate further into the bulk with increasing compression. In Fig. 4(c) and Fig. 4(d) we show beams emerging from the output facet of the array for $k_x = 0$ and $s = 1$ (where there is an edge state), and $k_x = 0$ and $s = 0.5$ (where there is none). The white ellipse indicates the position of the input beam. In the former case, light is confined to the edge due to the presence of the edge state, and in the latter case it diffracts into the bulk due to the lack thereof. Thus, we observe the complete destruction of the edge state associated with the bearded edge as a result of the merging of the Dirac points due to compression.

Numerical and experimental results for the zig-zag edge are shown in Fig. 5(a) and 5(b), respectively. This figure is the equivalent of Fig. 4, but for the zig-zag edge. Recall that the zig-zag edge exhibits topological edge states at exactly those values of $k_x$ for which the bearded does not. As the compression is increased, zig-zag edge states thus occupy all values of $k_x$. The topological diagram demonstrating the existence of an edge state throughout the edge Brillouin zone, equivalent to that of Fig. 2, is presented in the Appendix. Thus, just as the edge state was destroyed on the bearded edge as a function of increasing compression, it is created on the zig-zag edge. Beams emerging from the output facets of the waveguide arrays are shown in

Fig. 5(c) and Fig. 5(d) for $k_x = 0$ with $s = 1$ (where there is no edge state), and $s = 0.5$ (where there is an edge state), respectively. Fig. 4, the position of the input beam is indicated by the white ellipse in both Fig. 5(c) and Fig. 5(d). Clearly, no edge confinement is observed in the former case, and strong edge confinement is observed in the latter.

In summary, we have demonstrated theoretically and experimentally the destruction of graphene-like edge states on the bearded edge of a honeycomb photonic lattice, together with the simultaneous creation of edge states on the zig-zag edge. This result derives from a topological mechanism in which Dirac points merge and annihilate one another as a result of uniaxial lattice strain (compression). Topological systems are heralded for being impervious to small perturbations; this result is a prime example of how a large perturbation can tune between one topological phase and another.

Acknowledgements

M.C.R. is grateful for the support of a Fine Fellowship while at the Technion. A.S. gratefully acknowledges the support of the German Ministry of Education and Research (Center for Innovation Competence program, grant 03Z1HN31). M.S. gratefully acknowledges the support of the Israel Science Foundation, the USA-Israel Binational Science Foundation, and the Advanced Grant by the European Research Council.

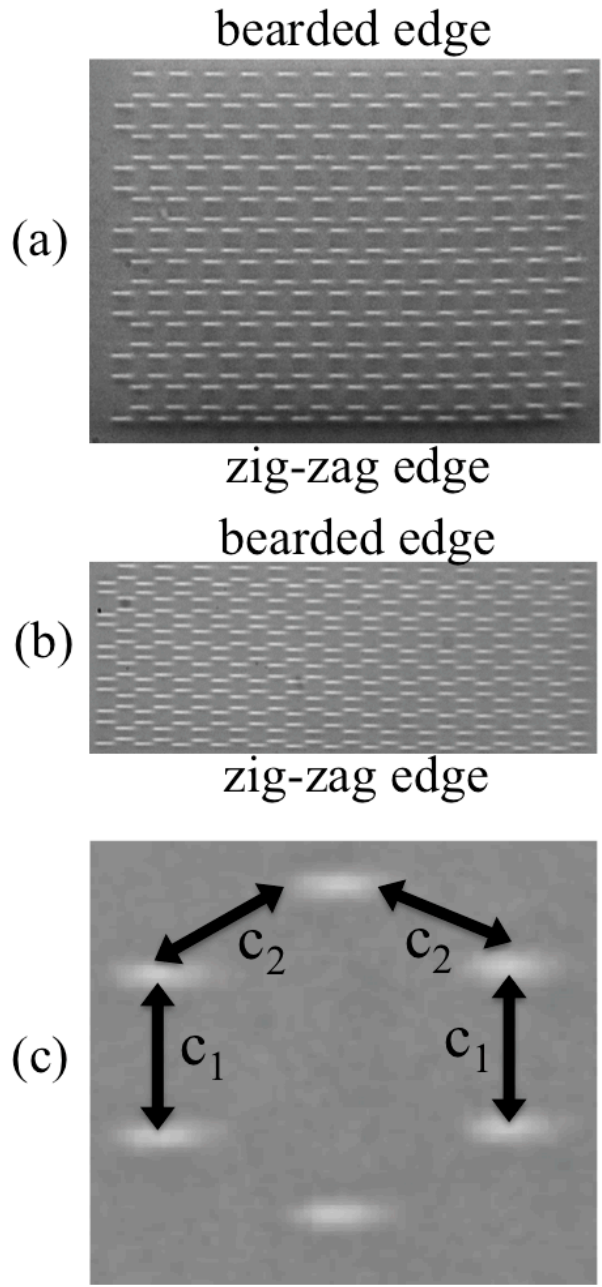

Fig. 1: Microscope images of the input facets of the uncompressed (a) and strongly compressed (b) honeycomb waveguide array. (c) The vertical coupling constant, $c_1$, and diagonal coupling constant, $c_2$.

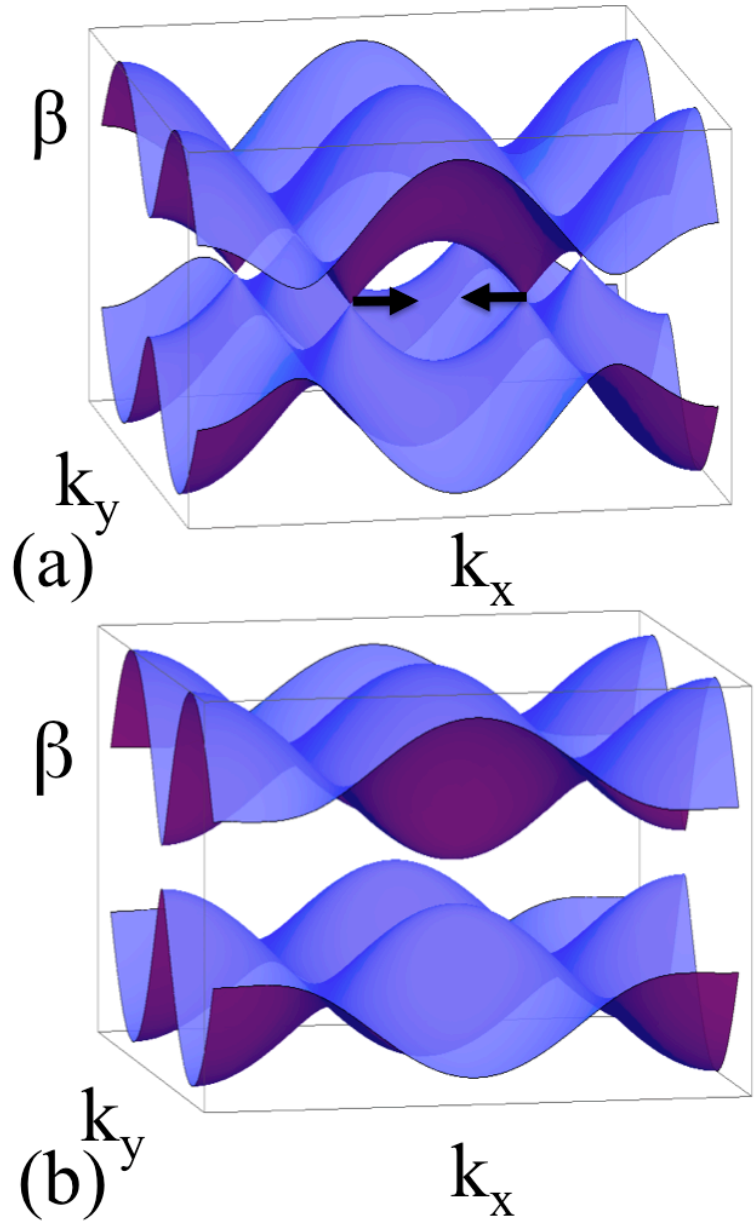

Fig. 2: Spatial band structure of the (a) uncompressed and (b) compressed honeycomb photonic lattice. The black arrows in (a) indicate the direction that the Dirac points move when the system is compressed.

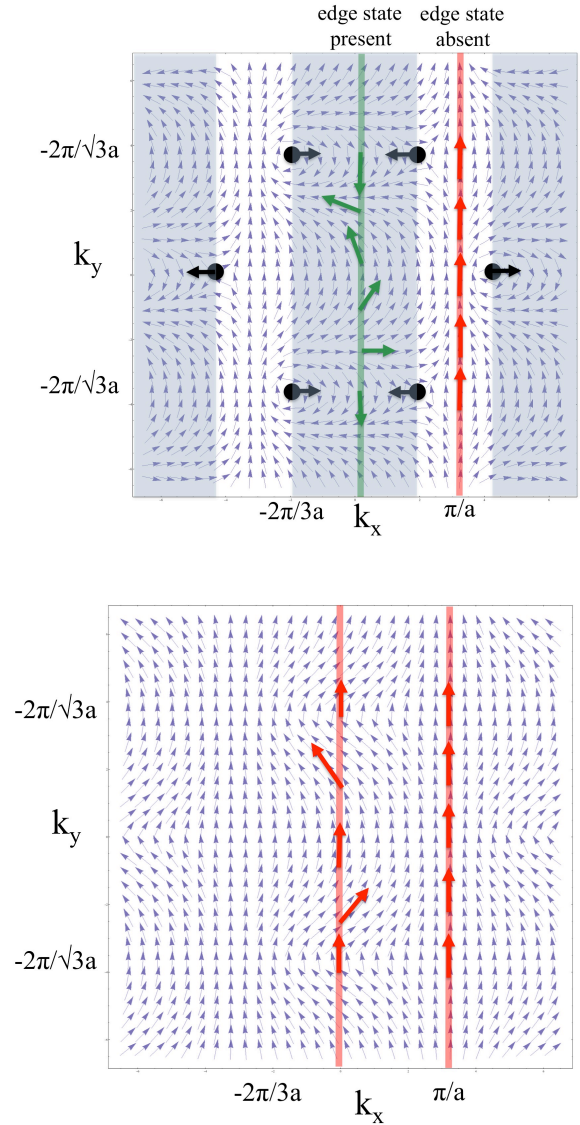

Fig. 3: Schematic of Berry's phase calculation demonstrating the values of $k_x$ for which an edge state exists, via the bulk-edge correspondence theorem, for both (a) the uncompressed ($s = 1$), and (b) strongly compressed ($s = 0.5$) cases. The arrows point in the direction of the vector $\mathbf{h}(\mathbf{k})$. In (a), the black dots indicate the Dirac points, and the arrows indicate the direction they move upon compression. Edge states are present in the shaded regions of (a), and are not present in (b).

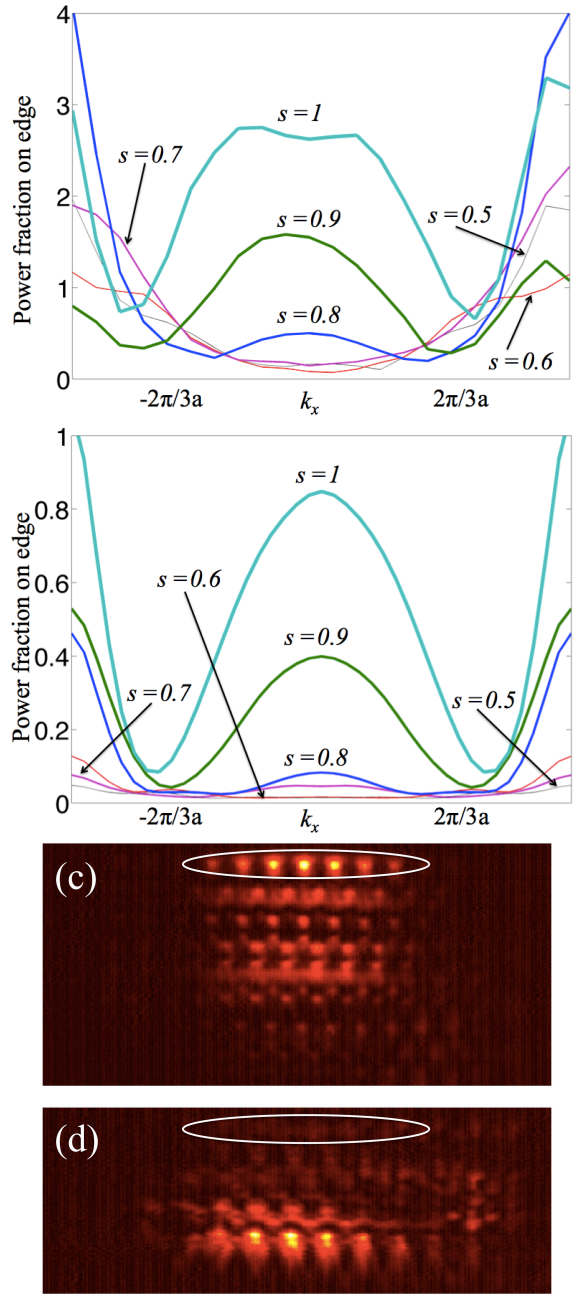

Fig. 4: Ratio of optical power on the bearded edge to that diffracted into the bulk, as a function of $k_x$: (a) shows experimental results, and (b) shows numerical results from beam-propagation simulations. (c) and (d) show the output beam at $k_x = 0$ when $s = 1$ and $0.5$, respectively. The white ellipse indicates the position of the input beam on the bearded edge.

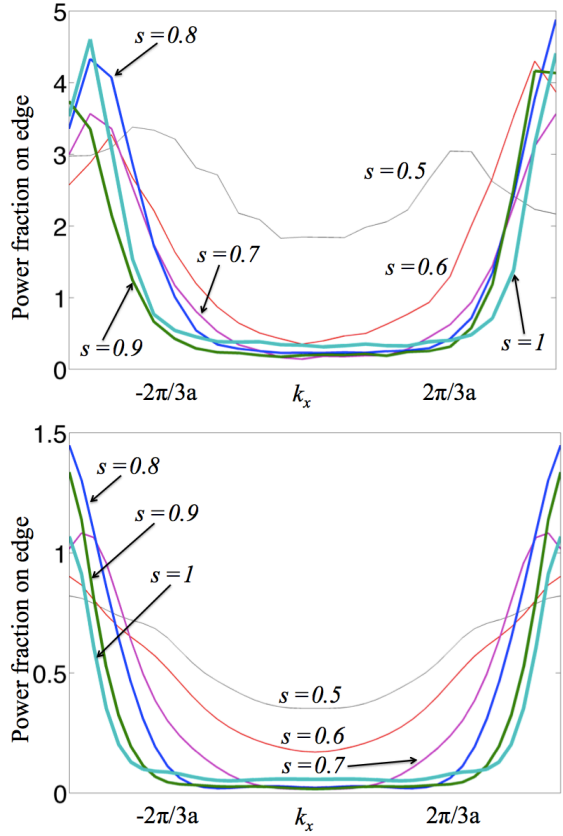

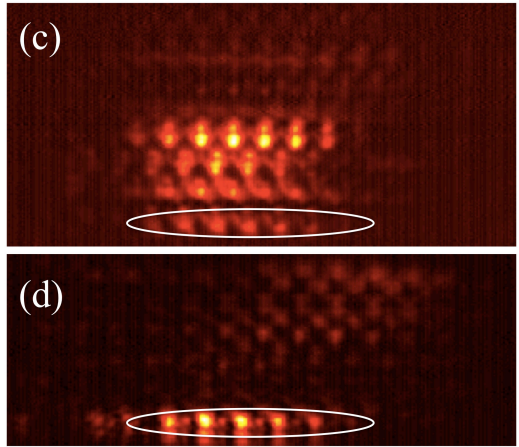

Fig. 5: Similar to Fig. 4, but for the zig-zag edge. Ratio of optical power on the zig-zag edge to that diffracted into the bulk, as a function of $k_x$: (a) shows experimental results, and (b) shows numerical results from beam-propagation simulations. (c) and (d) show the output beam at $k_x = 0$ when $s = 1$ and $0.5$, respectively. The white ellipse indicates the position of the input beam on the zig-zag edge.

# Appendix

Section 1: Berry phase of the Dirac cones

As discussed in the Letter, the Dirac points are 'topologically protected' against any perturbation that respects inversion symmetry, including the uniaxial strain, or compression, of the lattice. This means that small changes in the degree of compression cannot act to remove the conical touching of the Dirac points: a band gap cannot be opened. The reason for this has been explained previously [12,17], and the argument proceeds as follows. If the loop integral in **k**-space given in Eq. (2) is taken to surround a Dirac point, the result will be $\gamma = \pm\pi$, and in fact one unique Dirac point will have $\gamma = \pi$, and the other will have $\gamma = -\pi$, as long as inversion symmetry is preserved. Small compressions can change the positions of the Dirac points, but the phases cannot be changed and so the Dirac points cannot be removed. This is precisely what is meant by the topological protection of the Dirac points. If the compression is sufficiently strong, then the Dirac points can merge and the 'vorticities' associated with their respective Berry phases can thus cancel. It is only at this point that a band gap opens in the spectrum, as shown in Fig. 2(b) [17]. In Fig. S1, we show the graphical interpretation of Berry phase at the two unique Dirac points. As described in the Letter, if the vector **h**(**k**) makes a complete loop over the course of the integration path, the Berry phase will be $\gamma=\pm\pi$, otherwise it will be zero. Clearly, at both Dirac points, **h**(**k**) does indeed make a complete loop, but in opposite senses, indicating opposite signs of the Berry phase.

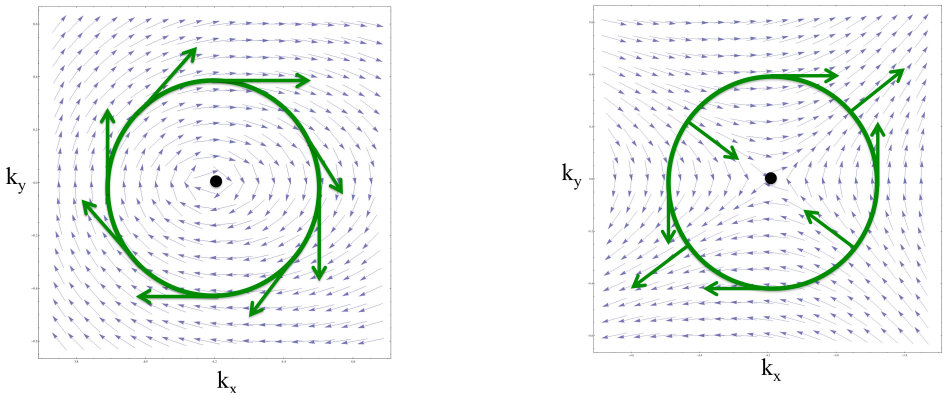

Fig. S1: Pictorial representation of the two Dirac points in the two-dimensional bulk Brillouin zone. Arrows represent the vector **h**(**k**). Clearly, **h**(**k**) makes a complete loop over the course of the circular integration path implying a Berry phase of $\pi$ for the clock-wise rotation (left), and $-\pi$ for the counter-clock-wise rotation (right).

Section 2: Bulk-edge correspondence diagram for zig-zag edge

In this section we show, using the bulk-edge correspondence theorem, which values of $k_x$ have edge states for the zig-zag edge. Note that this is exactly analogous to the discussion of Fig. 3 within the text (which was for the bearded edge). For the uncompressed lattice, the zig-zag edge as an edge state for $2\pi/3a < |k_x| < \pi/a$, within the first edge Brillouin zone (which occupies $|k_x| < \pi/a$). For the compressed lattice, the regions in which there is an edge state merge, and finally the zig-zag edge state occupies the entire edge Brillouin zone. In Fig. S2, in complete analogy with Fig. 3, we show a vector map of the direction of **h(k)** throughout the entire two-dimensional Brillouin zone (uncompressed in (a) - $c_1/c_2 = 1$ - and strongly compressed past the transition in (b) - $c_1/c_2 = 2.5$). Following the description in the main section of the Letter, the vertical lines that allow **h(k)** to trace out a full circle give rise to an edge state (indicated by the shaded regions). In the uncompressed case (Fig. S2(a)), this occurs for $2\pi/3a < |k_x| < \pi/a$; in the compressed case (Fig. S2(b)), this occurs for all $k_x$, indicating the presence of an edge state everywhere.

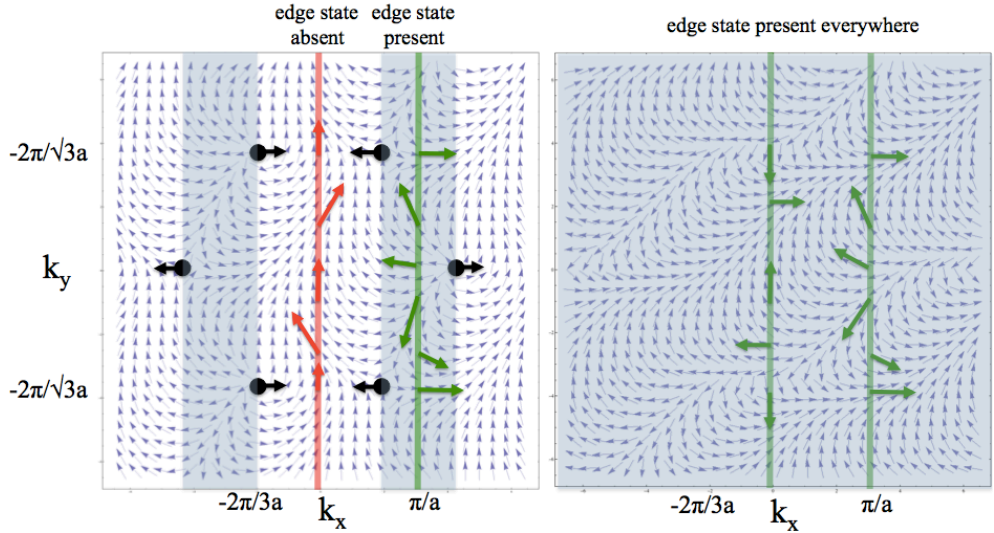

Fig S2: Pictorial representation of Berry's phase calculation indicating the presence of edge states on the zig-zag edge (analogous to Fig. 3 in the Letter). The bulk-edge correspondence principle states that when the arrows (indicating **h(k)**) form a complete loop along a vertical line, an edge state will be present at the corresponding $k_x$.

Section 3: Edge band structures for bearded and zig-zag edges

In order to explicitly demonstrate the destruction and creation of edge states as discussed in the text, we calculate the band structure of the honeycomb photonic lattice. We use a unit cell that is periodic in the *x*-direction but is terminated in the *y*-direction (both top and bottom) with both the bearded or zig-zag edge. The calculation is performed with the fully continuous Schrödinger equation (as opposed to the tight-binding model of Eq. (1)), and is thus an exact description of the spatial photonic band structure, with the experimental parameters given in the Letter. For each the bearded and zig-zag edges, the band structure calculation is shown for three values of the compression: $s = 1.0$ (uncompressed), $s = 0.8$ (slightly compressed), and s = 0.5 (fully compressed and after the merging of Dirac points). The band structures for each of these are shown in Fig. S3. The band structure gives the propagation constant, β, of a given eigenmode of the Hamiltonian (which is analogous to the negative of the energy in electronic systems obeying the Schrödinger equation). For the bearded edge, the fraction of the edge Brillouin zone that the edge state occupies shrinks as the system is compressed, and then disappears after the Dirac points merge. For the zig-zag edge, the opposite occurs: the edge state occupies a larger fraction of the edge Brillouin zone as the system is compressed. When the Dirac points merge, a zig-zag edge state occupies the entire edge Brillouin zone.

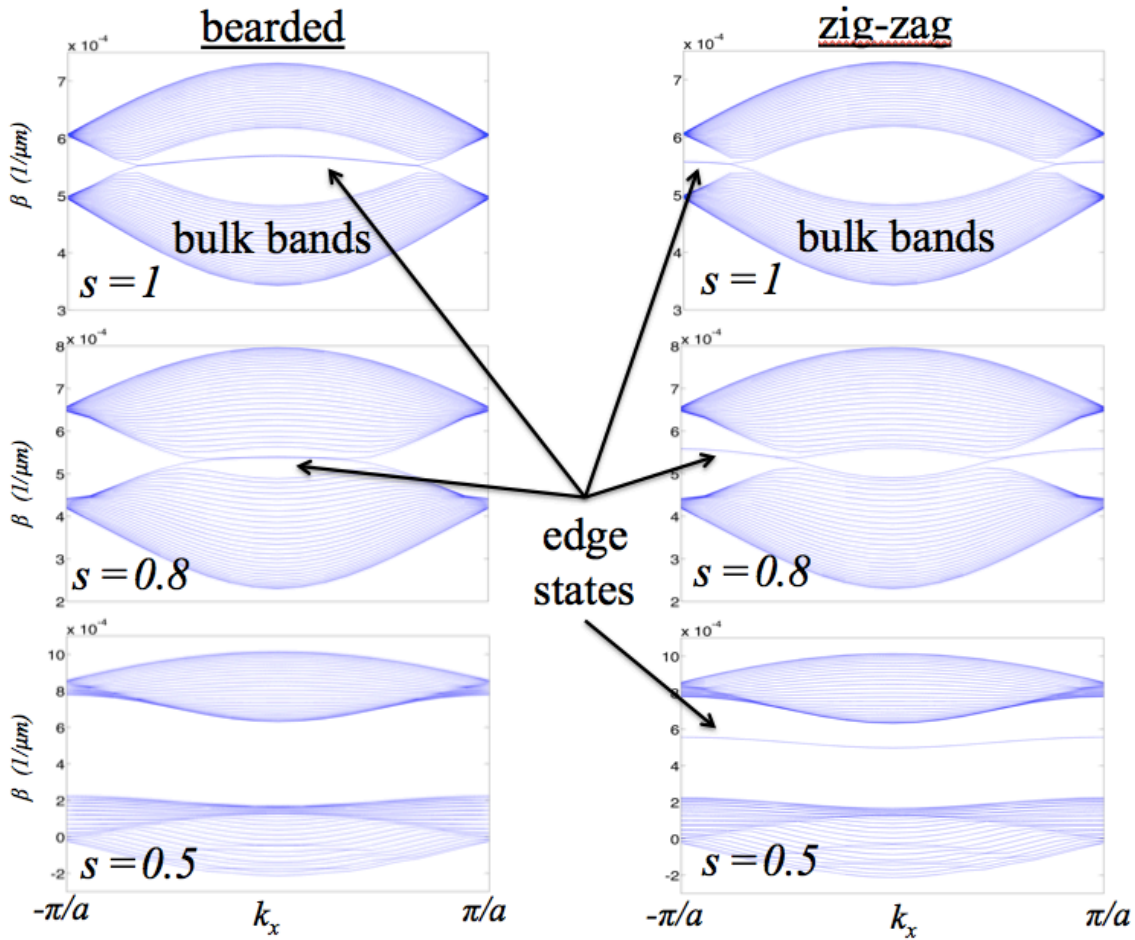

Fig. S3: Band structure diagrams: propagation constant, β, vs. $k_x$ for lattice that are terminated with the bearded edge (left column) and zig-zag edge (right column). In the top row, there is no lattice compression (s = 1), then mild, sub-critical compression (s = 0.8), and finally super-critical compression (s = 0.5), past the point where the Dirac points have merged and a band gap has opened.